\documentclass[twocolumn]{autart}    

\usepackage{hyperref}
\usepackage{textcomp}
\usepackage{caption}
\usepackage{algorithm}
\usepackage{amsmath,amssymb,amsfonts}
\usepackage{graphicx}          
\newcommand{\norm}[1]{\left\lVert#1\right\rVert}
\DeclareMathOperator*{\argmin}{arg\,min}
\newcommand*{\defeq}{\stackrel{\text{def}}{=}}
\newtheorem{theorem}{Theorem}[section]
\newtheorem{corollary}[theorem]{Corollary}

\usepackage{color}

\let\classAND\AND
\let\AND\relax
\usepackage{algorithmic}

\let\AND\classAND
\AtBeginEnvironment{algorithmic}{\let\AND\algoAND}

\begin{document}
\begin{frontmatter}

\title{Online Learning of Nonlinear Parametric Models under Non-smooth Regularization using EKF and ADMM}

\author[ODYS]{Lapo Frascati}\ead{lapo.frascati@odys.it},    
\author[IMT]{Alberto Bemporad}\ead{alberto.bemporad@imtlucca.it}  

\address[ODYS]{ODYS S.r.l., 20159, Milan, Italy}             
\address[IMT]{IMT School for Advanced Studies Lucca, 55100, Lucca, Italy}        

\thanks[footnoteinfo]{
The research work of Lapo Frascati has been financially supported by ODYS S.r.l. The research work of Alberto Bemporad has been funded by the European Union (ERC Advanced Research Grant COMPACT, No. 101141351). Views and opinions expressed are however those of the authors only and do not necessarily reflect those of the European Union or the European Research Council. Neither the European Union nor the granting authority can be held responsible for them. }
          

\begin{abstract}                          
This paper proposes a novel combination of extended Kalman filtering (EKF) with the alternating direction method of multipliers (ADMM) for 
learning parametric nonlinear models online under non-smooth regularization terms, including $\ell_1$ and $\ell_0$ penalties and bound constraints on model parameters. For the case of linear time-varying models and non-smooth convex regularization terms, 
we provide a sublinear regret bound that ensures the proper behavior of the online learning strategy. The approach is computationally efficient for a wide range of regularization terms, which makes it appealing for its use in embedded control applications for online model adaptation. We show the performance of the proposed method in three simulation examples, highlighting its effectiveness compared to other batch and online algorithms. 
\end{abstract}

\end{frontmatter}

\section{Introduction}
\label{sec:introduction}
Online learning of nonlinear parametric models is of paramount importance in several domains, including model-based adaptive control and real-time estimation of unmeasured variables. Typically, parametric models derived from physics~\cite{bib9} or black-box~\cite{bib8} structures are identified offline on training data, then directly deployed and used without any further updates. On the other hand, further adapting the model online can significantly improve its predictive capabilities~\cite{bib10}, especially when the phenomenon we are modeling changes over time, and allows for smaller model structures that adapt to varying operating conditions, unlike single, overall models trained offline to cover all conditions. 

A vast literature exists for online learning \cite{bib7}, and several approaches using first or second order gradient information are suitable for real-time model adaptation. Examples of algorithms exploiting first-order information are online gradient descent (OGD) and its regularized alternatives, i.e., follow the regularized leader (FTRL) and online mirror descent (OMD)~\cite{bib20}, which are usually fast at execution but slow at convergence. Second-order gradient information can be used to improve the convergence speed, such as in the online Newton step (ONS), or in the well known extended Kalman filter (EKF)~\cite{bib14} which has been proved to be very effective in online model adaptation by treating the parameters as constant states to be estimated~\cite{bib1,bib3}.

While FTRL and OMD, as well as modified versions of the standard EKF~\cite{bib1}, can effectively deal with {\it smooth} regularization terms, {\it non-smooth regularizers} are often required when learning models.
Non-smooth $\ell_0$ or $\ell_1$ penalties are used to induce sparsity in the model~\cite{bib25}, and group-Lasso terms to remove entire parts of the model, such as neurons of a neural network~\cite{bib26}. Indeed, obtaining compact models is
particularly important for embedded applications, such as model predictive control, 
where the real-time numerical complexity of the controller depends directly on the complexity of the prediction model. In addition, indicator functions of feasible sets are another example of non-smooth penalties that are used to impose constraints on model coefficients, such as known bounds on certain unknown physical parameters~\cite{bib5}.

Several approaches have been proposed for online learning under non-smooth regularization. Online ADMM (alternating direction method of multipliers)~\cite{bib2} allows handling quite general non-smooth regularizers, while EKF can be modified to deal with $\ell_1$-regularization by treating it as a special limit case of a smooth regularizer~\cite{bib1}. While the main limitation of online ADMM is its convergence speed, the main limitation of EKF is that it is not directly suitable for dealing with general non-smooth regularization terms, such as $\ell_0$ regularization, group-Lasso penalties, and indicator functions of feasible sets. The main contribution of this work is to show interesting connections and similarities between these two approaches, developing an extension of EKF that inherits the advantages from both, i.e., the convergence speed of EKF and the capability to handle general non-smooth regularization terms of online-ADMM.

The proposed method consists of a simple and computationally efficient modification of the EKF algorithm by intertwining updates based on online measurements and output prediction errors with updates related to ADMM iterations. This modification allows EKF to deal with a broad class of non-smooth regularization terms
for which ADMM is applicable, including $\ell_0$/$\ell_1$ penalties, group-Lasso and indicator functions of simple sets. 
For linear time-varying models and convex regularization terms, we provide a sublinear regret bound that proves the proper behavior of the resulting online learning strategy. The proposed method is computationally efficient and numerically robust, making it especially appealing for embedded adaptive control applications. 

The rest of the paper is organized as follows. Section~\ref{sec:EKF} gives a quick introduction to the use of EKF for online model learning, setting the background for the proposed ADMM+EKF approach described in Section~\ref{sec:ADMM}. In Section~\ref{sec:regret}, we prove a sublinear regret bound for the proposed approach in the convex linear case. Simulation results are shown in Section~\ref{sec:results} and conclusions are drawn in Section~\ref{sec:conclusions}. 

\subsection{Notation}
Given a matrix $A \in \mathbb{R}^{m\times n}$, $A_{i:}$ denotes the ith row of A, $A_{:j}$  its $j$th column, $A_{ij}$ its $(i,j)$th entry. Given a vector $v \in \mathbb{R}^n$ we denote by $v^m$ a measurement of $v$, by $\lVert v \rVert_1  = \sum_{i=1}^n | v_i |$ the 1-norm of $v$ and by $\lVert v \rVert_0$ its 0-norm, which is defined as the number of non-zero elements in the vector. Given a symmetric positive semidefinite matrix $P = P^T \succeq 0$, $P \in \mathbb{R}^{n\times n}$, we denote by $\lVert v \rVert^2_{P}$ the quadratic form $v^TPv$. Further, we denote by $\hat v_{i|j}$ the estimate of vector $v$ at instant $i$ given all information up to instant $j$, and by $P_{i|j}$ the corresponding covariance matrix. $v \sim \mathcal{N}(\mu,P)$ and $v \sim \mathcal{U}_{[v_{min},v_{max}]}$ denote that v was randomly generated from a normal distribution with mean $\mu$ and covariance $P$ or from a uniform distribution in the interval $[v_{min},v_{max}]$, respectively.

\section{EKF for online model learning}
\label{sec:EKF}
Given a dataset $(z_k^m,y_k^m)$, $z\in\mathbb{R}^{n_z}$,
$y\in\mathbb{R}^{n_y}$, $k=0,1,\ldots, N-1$, our goal is to recursively estimate a nonlinear parametric model 
\begin{equation}
    y = h(k,z;x)
    \label{eq:static_model}
\end{equation}
which describes the (possibly time-varying) relationship between the input signal $z_k^m$ and the output signal $y_k^m$. In~\eqref{eq:static_model}, $x\in \mathbb{R}^{n_x}$ is the vector of parameters to be learned, such as the weights of a feedforward neural network mapping $z$ into $y$, or the coefficients of a nonlinear autoregressive model, with $y$ representing the current output and $z$ a vector of past inputs and outputs of a dynamical system. In order to estimate $x$ and capture its possible time-varying nature, we consider the nonlinear dynamical model
\begin{equation}
x_{k+1} = x_{k} + q_k,\quad y_k^{nl} = h_k(x_k) + r_k
\label{ekfmodel0}
\end{equation}
where 
$h_k(\cdot)=h(k,z_k;\cdot)$ and we assume
$h_k: \mathbb{R}^{n_x} \rightarrow \mathbb{R}^{n_y}$ differentiable for all $k$, $y_k^{nl} \in \mathbb{R}^{n_y}$, and $r_k \sim \mathcal{N}(0,R_k), q_k \sim \mathcal{N}(0,Q_k)$ are the measurement and process noise that we introduce to model, respectively, measurement errors and variations of the model parameters over time,
with corresponding covariance matrices $R_k=R_k'\succ 0$, $Q_k=Q_k'\succ 0$. By linearizing model~\eqref{ekfmodel0} around a value $\overline x_k$
of the parameter vector, i.e., by approximating $h_k(x_k)\approx h_k(\overline x_k) + C_k (x_k-\overline x_k)$, $C_{k,i:} = \nabla_x h_{ki}(\overline x_k)'$, $i=1,\ldots,n_y$,
we obtain the linear time-varying model
\begin{equation}
x_{k+1} = x_{k} + q_k,\quad  y_k = C_k x_k + r_k
\label{ekfmodel}
\end{equation}
with $y_k = y_k^{nl} -h_k(\overline x_k) + C_k \overline x_k$. The classical Kalman filter \cite{bib14} can be used to estimate the state in \eqref{ekfmodel}, i.e., to learn the parameters $x_k$ recursively. Given $\hat x_{0|-1}, P_{0|-1}$ we perform the following iterations for $k = 0, \dots, N-1$:
\begin{equation}
\begin{aligned}
& P_{k|k}^{-1} = P_{k|k-1}^{-1}+C_k^TR_k^{-1}C_k \\
& \hat x_{k|k} = \hat x_{k|k-1} + P_{k|k}C_k^TR_k^{-1} (y_k^m - C_k \hat x_{k|k-1}) \\
& P_{k+1|k}^{-1} = (Q_{k}+P_{k|k})^{-1} \\
& \hat x_{k+1|k} = \hat x_{k|k} \\
\end{aligned}
\label{kfilter}
\end{equation}
with $\overline x_k = \hat x_{k|k-1}$. The first two updates in \eqref{kfilter} are usually referred to as the correction step and the last two as the prediction step. 
Note that~\eqref{kfilter} is an EKF for model~\eqref{ekfmodel0}, since $C_k$ is the Jacobian of the output function at $\hat x_{k|k-1}$ and the output prediction error
used in the correction step is $e_k =y_k^m - C_k \hat x_{k|k-1} = y_k^{nl,m} -h_k(\hat x_{k|k-1})$. 

As shown in \cite{bib13}, the state estimates $\hat x_{k|k}, \hat x_{k+1|k}$ 
generated by the Kalman filter~\eqref{kfilter} are part of the optimizer of the following optimization problem
\begin{equation}
\begin{aligned}
& \hat x_{0|k}, \dots, \hat x_{k|k}, \hat x_{k+1|k} = \argmin_{x_{0}, \dots, x_{k}, x_{k+1}} \norm{x_{0} - \hat x_{0|-1}}^2_{P_{0|-1}^{-1}} + \\
& + \sum_{i=0}^{k}\norm{y_{i}^m - C_{i}x_{i}}^2_{R_{i}^{-1}} + \norm{x_{i+1} - x_{i}}^2_{Q_i^{-1}}.
\label{completecost}
\end{aligned}
\end{equation}
Problem~\eqref{completecost} can be solved recursively at each step $k$ by minimizing the following cost functions:
\begin{subequations}
\begin{eqnarray}
 &&\hat x_{k|k} = \argmin_{x_{k}}\norm{x_{k} - \hat x_{k|k-1}}^2_{P_{k|k-1}^{-1}}\hspace*{-1em} + \norm{y_{k}^m - C_{k}x_{k}}^2_{R_{k}^{-1}}
\nonumber\\~\label{reducedform2}\\[-1em]
&&\hat x_{k|k}, \hat x_{k+1|k} = \argmin_{x_k,x_{k+1}}\norm{x_{k} - \hat x_{k|k}}^2_{P_{k|k}^{-1}} +\nonumber\\
&&\hspace*{12em}  + \norm{x_{k+1} - x_{k}}^2_{Q_{k}^{-1}}
\label{reducedform3}
\end{eqnarray}
\end{subequations}
where $\hat x_{k|k}, \hat x_{k+1|k}, P_{k|k}$, and $P_{k+1|k}$ are the state estimates and covariance matrices computed as in \eqref{kfilter}.

\section{EKF under non-smooth regularization}
\label{sec:ADMM}
In order to regularize the model, we modify the classical iterations \eqref{kfilter} by changing the minimization in~\eqref{reducedform2} to
\begin{equation}
\begin{aligned}
 \hat x_{k|k} = &\argmin_{x_{k}}\frac{1}{2}\norm{x_{k} - \hat x_{k|k-1}}^2_{P_{k|k-1}^{-1}} + \\
& + \frac{1}{2}\norm{y_{k}^m - C_{k}x_{k}}^2_{R_{k}^{-1}} + g(x_k)
\label{eacost1}
\end{aligned}
\end{equation}
where $g(\cdot): \mathbb{R}^{n_x} \rightarrow \mathbb{R} \cup \{ + \infty\}$ is a possibly non-smooth and non-convex regularization term. By defining $\mathcal{S} = \{(x_k,\nu)\in\mathbb{R}^{n_x}\times\mathbb{R}^{n_x}:\  x_k=\nu\}$,~\eqref{eacost1} can be equivalently reformulated as the following constrained optimization problem
\begin{equation}
\begin{aligned}
 \hat x_{k|k}, \nu^\star = & \argmin_{(x_k,\nu)\in \mathcal{S}}\frac{1}{2}\norm{x_{k} - \hat x_{k|k-1}}^2_{P_{k|k-1}^{-1}} + \\
& + \frac{1}{2}\norm{y_{k}^m - C_{k}x_{k}}^2_{R_{k}^{-1}} + g(\nu)
\label{eacost2}
\end{aligned}
\end{equation}
which can be solved by executing the following scaled ADMM iterations \cite{bib5}:
\begin{subequations}
\begin{eqnarray}
 \hat x_{k|k}^{t+1} &=& \argmin_{x_{k}}\norm{x_{k} - \hat x_{k|k-1}}^2_{P_{k|k-1}^{-1}} +\nonumber \\
&& + \norm{y_{k}^m - C_{k}x_{k}}^2_{R_{k}^{-1}} + \rho \norm{x_k - \nu^t + w^t}_2^2
\label{admm1}
\\
\nu^{t+1}& = & \argmin_{\nu}g(\nu)+\frac{\rho}{2}\norm{\nu - \hat x_{k|k}^{t+1} - w^{l}}_2^2\nonumber\\
& =&  \text{prox}_{\frac{g}{\rho}}(\hat x_{k|k}^{t+1} + w^t)
\label{admm2}
\\
w^{t+1} &= &w^{t} +\hat x_{k|k}^{t+1} - \nu^{t+1}
\label{admm3}
\end{eqnarray}
\end{subequations}
for $t = 0, \dots, n_{a}-1$, where $\rho>0$ is a hyper-parameter to
be calibrated and ``prox'' is the proximal operator~\cite{bib15}. As shown in \cite{bib5}, in the convex case, the ADMM iterations \eqref{admm1}--\eqref{admm3} converge to the optimizer of \eqref{eacost2} as $n_a \rightarrow \infty$, and often converge to a solution 
of acceptable accuracy within a few tens of iterations. Iteration \eqref{admm3} is straightforward to compute; iteration \eqref{admm2} can be solved explicitly and efficiently with complexity $\mathcal{O}(n_x)$ for a wide range of non-smooth and non-convex regularization functions $g$, such as $g(x)=\|x\|_0$, $g(x)=\|x\|_1$,
and the indicator function $g(x)=0$ if $x_{\rm min}\leq x\leq x_{\rm max}$ or $+\infty$ otherwise~\cite{bib15}. Iteration \eqref{admm1} can be rewritten as
\begin{equation}
\begin{aligned}
& \hat x_{k|k}^{t+1} = \argmin_{x_{k}}\norm{x_{k} - \hat x_{k|k-1}}^2_{P_{k|k-1}^{-1}} +
\norm{\overline{y}_k^m - \overline{C}_k x_{k}}^2_{\overline R_k^{-1}}
\label{admm1-mod}
\end{aligned}
\end{equation}
where $\overline{y}_k^m = [(y_{k}^m)'\ (\nu^t-w^t)']'$, $\overline{C}_k = [C_{k}'\ I]'$, and $\overline{R}_k = \left[\begin{smallmatrix}R_{k} & 0 \\ 0 & \rho^{-1}I\end{smallmatrix}\right]$. Therefore, iteration \eqref{admm1} can be performed directly in the correction step of the EKF by including $n_{x}$ additional ``fake'' state measurements $\nu^t-w^t$ with covariance matrix $\rho^{-1} I$. 

Algorithm \ref{alg1} summarizes the proposed extension of EKF with ADMM iterations (EKF-ADMM). The algorithm returns the estimate $\hat x_{k|k}$ of the parameter vector $x$ obtained after processing $N$ measurements. It also returns the last value of $\nu$, which could be used as an alternative estimate of $x$ too; for example, in case $g$ is the indicator function of a constraint set, $\nu$ would be guaranteed to be feasible. Note that the dual vector $w$ is not reset at each EKF iteration $k$; it is used as a warm start for the next $n_a$ ADMM iterations at step $k+1$, as the solutions $\hat x_{k|k}$ at consecutive time instants $k$ are usually similar.
\subsection{Computational complexity}
Given the block-diagonal structure of the measurement noise covariance matrix $\overline R_k$,~\eqref{admm1-mod} can be rewritten as 
\begin{equation}
\begin{aligned}
 \hat x_{k|k}^{t+1} & = \argmin_{x_{k}}\norm{x_{k} - \hat x_{k|k-1}}^2_{P_{k|k-1}^{-1}} + \\
& + \norm{y_{k}^m - C_{k}x_{k}}^2_{R_{k}^{-1}} + \rho \sum_{i=1}^{n_x}\norm{x_{k,i} - \nu^t_i + w^t_i}_2^2
\end{aligned}
\end{equation}%
which highlights the separation of the contributions of the true measurements $y_k^m$ and of the fake regularization measurements $\nu^{t} - w^{t}$ in the correction step; moreover, we can process the measurements $\nu^{t}_i - w^{t}_i$ separately one by one. This allows designing a computationally more efficient and numerically robust version of the proposed EKF-ADMM algorithm, as the correction due to the true measurements $y_k^m$ can be performed only once, instead of $n_a$ times, as it does not change with $t$. Moreover, there is no need for any matrix inversion when processing the fake measurements, as by processing them one by one, the matrix inversion required to compute the Kalman gain becomes a simple division, since each measurement is just a scalar value. Assuming a complexity $\mathcal{O}(n_x)$  for evaluating the proximal operator, EKF-ADMM has complexity $\mathcal{O}(n_x^3 + n_{a}n_x^2)$, which is the same order of the full EKF for general state estimation. Moreover, EKF-ADMM has the same number of Jacobian matrices evaluations than the classical EKF, which is usually the most time-consuming part in case $x$ represents the weights and bias terms of a neural network model to learn. Summarizing, the proposed approach is computationally efficient and, if the Kalman filter is implemented using numerically robust factored or square-root modifications \cite{bib16}, the method is appealing for embedded applications.
\begin{algorithm}
\caption{EKF-ADMM}\label{alg:alg1}
\begin{algorithmic}
\REQUIRE $\hat x_{0|-1}, P_{0|-1}^{-1}, \nu = \hat x_{0|-1}, w  = 0, \rho>0 $ 
\FOR {$k = 0, \dots, N-1$}
\STATE $K_k = P_{k|k-1}\overline{C}_k^T(\overline{R}_k+\overline{C}_kP_{k|k-1}\overline{C}_k^T)^{-1}$
\FOR {$t = 0, \dots, n_{a}-1 $}
\STATE $\hat x_{k|k} \gets \hat x_{k|k-1}+K_k\bigg{(}\begin{pmatrix} y_{k}^m \\ \nu-w \end{pmatrix}- \overline{C}_{k}\hat x_{k|k-1}\bigg{)}$ 
\STATE $\nu \gets \text{prox}_{\frac{g}{\rho}}(\hat x_{k|k} + w)$
\STATE $w \gets w + \hat x_{k|k} - \nu$
\ENDFOR
\STATE $P_{k|k} = (I - K_k\overline{C}_k)P_{k|k-1}$
\STATE $\hat x_{k+1|k} = \hat x_{k|k}$
\STATE $P_{k+1|k} = P_{k|k} + Q_{k}$
\ENDFOR 
\RETURN $\hat x_{N-1|N-1}$, $\nu$
\end{algorithmic}
\label{alg1}
\end{algorithm}
\section{Regret analysis}
\label{sec:regret}
We investigate the theoretical properties of EKF-ADMM for linear time-varying models, i.e., models of the form 
\begin{equation}
    y_k = h_k(x_k) = C_k x_k
    \label{eq:LTV-model}
\end{equation}
where $C_k$ are now given time-varying matrices for $k=0,1,\ldots,N-1$,
and convex regularization terms $g$. In particular, we want to evaluate the ability of the algorithm to solve the optimization problem $\min_{x} \sum_{k=0}^{N-1} (f_k(x) + g(x))$ online, where $f_k(x)=\frac{1}{2}\norm{y_k^m - C_k x}^2_{R_{k}^{-1}}$, via the following two regret functions $R_{f}(N) = \sum_{k=0}^{N-1} (f_k(x_k)+g(\nu_k)) - \min_{x,\nu \in \mathcal{S}}\sum_{k=0}^{N-1} (f_k(x)+g(\nu))$ and $R_{c}(N) = \sum_{k=0}^{N-1} \norm{x_{k+1}-\nu_{k}}^2 $, where, to simplify the notation, we have defined $x_k = \hat x_{k|k-1}, P_k = P_{k|k-1}, \forall k=0,1,\ldots, N-1$. 
Notice that $R_{f}(N)$ quantifies the loss we suffer by learning the model online instead of solving
it in a batch way given all $N$ measurements, while $R_{c}(N)$ quantifies the violation
of the constraint $x=\nu$. To ensure a proper behavior of EKF-ADMM, we want to prove a sublinear regret bound for both, i.e., $R_{f}(N) \leq \mathcal{O}(\sqrt{N})$ and $R_{c}(N) \leq \mathcal{O}(\sqrt{N})$ \cite{bib2}. 

EKF-ADMM is a generalization of the online ADMM method proposed in \cite{bib2}, in which a sublinear regret bound is derived for the case $n_a=1$ and $P_{k}^{-1}=P^{-1}\succ0, \forall k$, while, more recently, in \cite{bib18} a sublinear regret bound has been derived for the case $n_a=1$ and $P_{k}^{-1}\succeq P_{k+1}^{-1}, \forall k$. Here we will provide a sublinear regret in the case $n_a=1$ and $P_{k}^{-1} = P_{k+1}^{-1}, \forall k\geq k_n \ll N$, which is a reasonable assumption as the EKF covariance matrix, when estimating the parameters of a model, usually has a transient and then reaches a steady-state value. 
By assuming $n_a=1$ and $P_{k}^{-1} = P_{k+1}^{-1}, \forall k \geq k_n \ll N$, Algorithm \ref{alg1} can be equivalently rewritten as in Algorithm \ref{alg3}.

\begin{algorithm}
	\caption{EKF-ADMM ($n_a=1$, frozen $P$)} \label{alg:alg3}
	\begin{algorithmic}
	 \REQUIRE{$x_{0},P_{0}^{-1},\nu_1= x_{0},w_0=0, \rho, \eta>0$, $k_n\geq 0$}
		\FOR {$k=0,\dots, N-1$}
		\STATE $\begin{aligned} x_{k+1} \gets &\argmin_{x} \frac{1}{2} \norm{y_{k}^m - C_{k}x}^2_{R_{k}^{-1}} + w_k^T (x - \nu_k) + \\
& + \frac{\rho}{2}\norm{x - \nu_k}^2 + \frac{\eta}{2} \norm{x -  x_{k}}^2_{P_{k}^{-1}}\end{aligned}$
		\STATE $\begin{aligned}\nu_{k+1} \gets &\argmin_{\nu}g(\nu)+w_k^T(x_{k+1}-\nu)+ \\
		& + \frac{\rho}{2}\norm{x_{k+1}-\nu}^2= \text{prox}_{\frac{g}{\rho}}(x_{k+1} + \frac{w_k}{\rho}) \end{aligned}$
		\STATE $w_{k+1} \gets w_k +\rho (x_{k+1} - \nu_{k+1})$ 
	    \IF{$k < k_n$}
		\STATE $P_{k+1}^{-1} \gets (Q_{k}+(P_{k}^{-1}+ \overline C_k^T \overline R_k^{-1}\overline C_k)^{-1})^{-1} $
		\ELSE
		\STATE $P_{k+1}^{-1} \gets P_{k}^{-1}$
	    \ENDIF
		\ENDFOR
        \RETURN $x_N,\nu_N$
	\end{algorithmic} 
	\label{alg3}
\end{algorithm}
The following Theorem~\ref{thm:regret} is an extension of \cite[Theorem 4]{bib2}, and provides conditions for sublinear regret bounds of Algorithm \ref{alg3} in the case of a linear time-varying model~\eqref{eq:LTV-model} and convex regularization function $g$.
\begin{theorem}
\label{thm:regret}
Let $\{x_k, \nu_k,w_k\}_{k=0}^{N-1}$ be the sequence generated by Algorithm  \ref{alg3} and let $x^{\star},\nu^{\star}$ be the best solution in hindsight, i.e. $ x^{\star},\nu^{\star} = \argmin _{x,\nu \in \mathcal{S}} \sum_{k=0}^{N-1} (f_k(x)+g(\nu))$. 
Let the following assumptions hold: 
\begin{enumerate}
\item [A1.] $\exists\alpha, G_f, D_x, D_\nu, F>0$ such that $\forall k = 0, \dots, N-1$:
\begin{enumerate}
\item $\norm{x-y}^2_{P_k^{-1}} \geq \alpha\norm{x-y}_2^2$, $\forall x,y$
\item $\norm{\nabla f_k(x_{k})}_2^2 = \norm{C_k^TR_k^{-1}(C_kx_{k}-y_k^{m})}_2^2 \leq G_f^2$ 
\item $\frac{1}{2}\norm{x^\star}^2_{P_{k}^{-1}} \leq D_x^2$ and $\norm{\nu^\star}^2_2\leq D_{\nu}$ 
\item $f_k(x_{k+1})+g(\nu_{k+1})-(f_k(x^\star) + g(\nu^\star))\geq -F$
\end{enumerate}
\item [A2.]$\exists M_{k_n}\geq0$ such that $\frac{1}{2}\sum_{k=1}^{k_n} \norm{x^\star - x_{k}}^2_{(P_{k}^{-1}-P_{k-1}^{-1})} \leq M_{k_n}$
\item [A3.]To ease the notation, $x_{0}=0, g(0)=0$ and $g(\nu) \geq 0$.
\end{enumerate}
Then, if $\eta = \frac{G_f \sqrt{N}}{D_x \sqrt{2\alpha}}$ and $\rho = \sqrt{N}$, the following sublinear regret bounds are guaranteed: 
\begin{subequations}
\begin{equation}
R_f(N)\leq \frac{\sqrt{N} D_{\nu}}{2} + \frac{G_f D_x\sqrt{N} }{\sqrt{2\alpha}} + \frac{G_f \sqrt{N}(D_x^2 + M_{k_n})}{D_x \sqrt{2\alpha}}
\end{equation}
\begin{equation}
R_c(N) \leq 2F\sqrt{N} + D_{\nu} + \frac{2G_{f}}{D_x \sqrt{2\alpha}} (D_x^2+M_{k_n}).
\end{equation}
\end{subequations}
\end{theorem}
\textbf{Proof.} See Appendix~\ref{app:Theorem1}.\hfill$\Box$

\begin{corollary}
Consider the linear time-invariant case $C_k\equiv C_0$, $\forall k\geq 0$. If the steady-state Kalman filter is used, then Theorem~\ref{thm:regret} holds with $M_{k_n} = 0$. \end{corollary}
In general, as proved in \cite{bib18}, Theorem~\ref{thm:regret} holds with $M_{k_n} = 0$ whenever $P_{k}^{-1}\succeq P_{k+1}^{-1}, \forall k$. Intuitively, this means that for online model adaptation we need to limit the importance of the previous samples to promptly adapt the model to changes and therefore bound the regret function. This can be accomplished, for example, using the EKF with a proper forgetting factor~\cite{bib3}. 

\section{Simulation results}
\label{sec:results}
We evaluate the performance of the proposed EKF-ADMM algorithm on three different examples: online LASSO~\cite{bib22}, online training of a neural network on data from a static model under $\ell_1$ regularization or bound constraints, online adaptation of a neural network on data from a time-varying model under $\ell_0$ regularization.

\subsection{Online LASSO}
\begin{figure}[!t]
\centerline{\includegraphics[width=0.95\columnwidth]{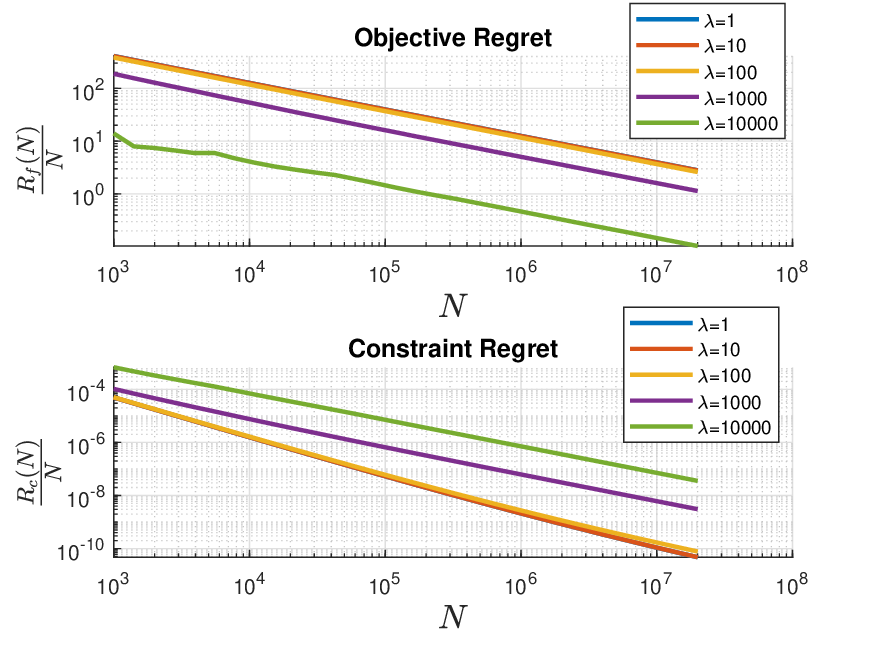}}
\caption{Objective and constraint regret for online LASSO.}
\label{regret_lasso}
\end{figure}

Consider the LASSO problem $\min_{x} \sum_{k=0}^{N-1}( \frac{1}{2}\left\|y_k^m\right.$ $-$ $\left.C_k x\right\|^2_{R^{-1}}$ $+$ $\lambda\norm{x}_1)$, where $x \in \mathbb{R}^{3}$ is the parameter vector, $C_k\in \mathbb{R}^{2 \times 3}$ are randomly generated matrices with $C_{k,ij} \sim \mathcal{N}(0,1)$, $y_k^m = C_k x_{\rm true} + r_{k} \in \mathbb{R}^2$ is the vector of measurements and $r_k$ is random measurement noise, $r_{k,i}\sim{\mathcal N}(0,10^{-2})$. We will evaluate the behavior of the regret functions $R_f(N)$ and $R_c(N)$ as $N \rightarrow \infty$ when using Algorithm \ref{alg3}. The following settings are used: $P_{0|-1} = I, Q_k = 10^{-6}I, R = 10^{-3}I, k_n = 10^3, \rho = 10^4\sqrt{N}$ and $\eta = 10^{-6}\sqrt{N}$, where $P_{0|-1}, Q_k$ and $R$ are manually tuned to maximize the Kalman filter performance without regularization and $\rho, \eta$ are chosen according to the expressions provided in Theorem~\ref{thm:regret}. Results for different values of $\lambda$ are shown in Figure~\ref{regret_lasso}. In this case, Theorem 2 holds and, as expected, both the regrets $R_f(N)$ and
$R_c(N)$ decrease as the number $N$ of samples increases.

\subsection{Online learning of a static model}
Consider the dataset generated by the static nonlinear model $y^m_k = \frac{z_{k,1}^2-e^{\frac{z_{k,2}}{10}}}{3+|z_{k,1}+z_{k,2}|} + r_k$. We want to train online a neural network $h_k(x)$ with $2$ layers, $8$ neurons in each layer,
and $\tanh$ activation function, with $n_x=105$ trainable weights in total. The training is performed on $N = 10^5$ randomly generated data points, where $z_{k,i} \sim \mathcal{U}_{[-10,10]}$ and $r_{k}\sim{\mathcal N}(0,10^{-2})$. Let $\{ x_k\}_{k=0}^{N-1}$ be the sequence of weights generated by Algorithm \ref{alg1}: we evaluate the online adaptation performance by means of the regret function $R_{f}(N) = \sum_{k=0}^{N-1} (f_k(x_k)+g(x_k)) - \min_{x}\sum_{k=0}^{N-1} (f_k(x)+g(x))$, where $f_k(x) = \frac{1}{2}\norm{y_k^m - h_k(x)}^2$, and the quality of a given solution $x$ using the performance indices $\text{Loss}(x) = \frac{1}{N}\sum_{k=0}^{N-1} (f_k(x)+g(x))$, $\text{Mse}(x) = \frac{1}{N}\sum_{k=0}^{N-1} f_k$, $\text{Reg}(x) = \frac{1}{N} \sum_{k=0}^{N-1} g(x)$ and $\text{Cv}(x) = \norm{x - \Pi_{\mathcal{C}}(x)}_2^2$, where, given a constraint set $\mathcal{C} \subseteq \mathbb{R}^{n_x}$, $\Pi_{\mathcal{C}}(x)$ is the projection of the point $x$ onto $\mathcal{C}$. The training is performed in MATLAB R2022a on an Intel Core i7 12700H CPU with 16 GB of RAM, using the library CasADi~\cite{AGHRD19} to compute the required Jacobian matrices via automatic differentiation. 
All results are averaged over 20 runs starting from different initial conditions, that were randomly generated using the well known Xavier's initialization strategy. 

\subsubsection{$\ell_1$ regularization}
\begin{figure}[!t]
\centerline{\includegraphics[width=0.95\columnwidth, trim=12 10 30 8, clip]{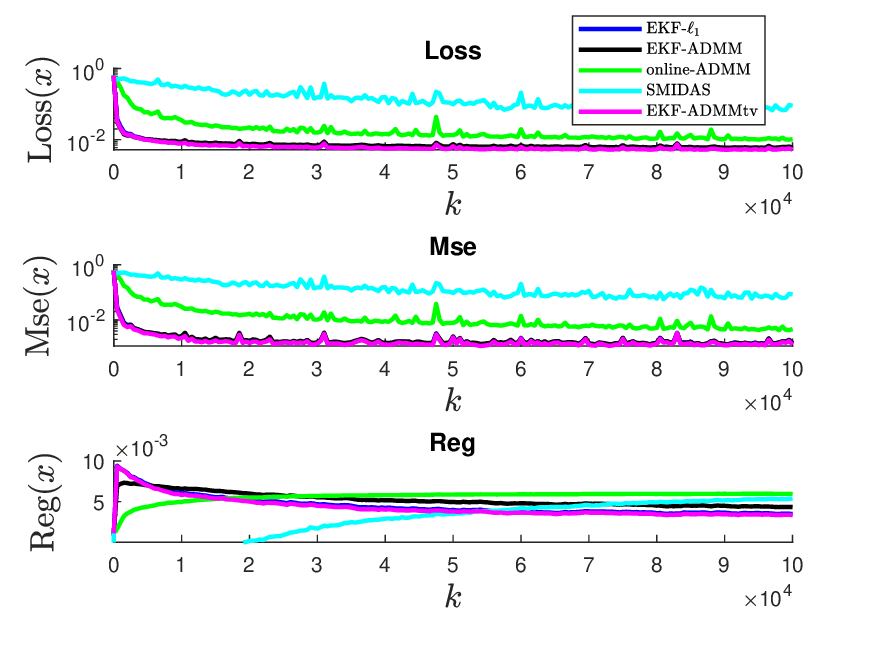}}
\caption{Online learning with $\ell_1$ regularization: Loss, Mse and Reg averaged over $20$ runs.}
\label{nn_static_l1_loss}
\end{figure}

\begin{figure}[!t]
\centerline{\includegraphics[width=0.94\columnwidth, trim=6 10 30 8, clip]{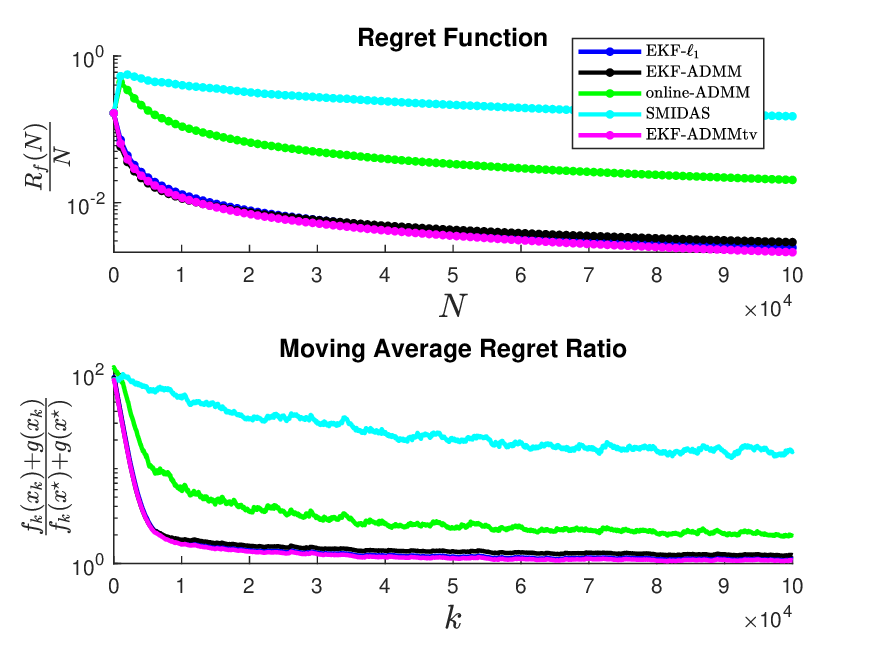}}
\caption{Online learning with $\ell_1$ regularization: regret and sample regret averaged over $20$ runs.}
\label{nn_static_l1_regret}
\end{figure}

We train the neural model under the regularization function $g(x) = \lambda \norm{x}_1$,
with $\lambda = 10^{-4}$. We selected the following hyper-parameters:  $\rho = 10 \lambda$, $n_a = 1$, $Q_k = 10^{-4} I$, $R_k = I$ and $P_{0|0} = 100 I$. We compare the results to different online optimization alternatives: online ADMM~\cite{bib2} with constant matrix $P = 10^{-2}I$ (online-ADMM), EKF-ADMM with time-varying $\rho_k = 10^{\frac{k}{N}-2} \lambda$ (EKF-ADMMtv), EKF with $\ell_1$-regularization~\cite{bib1} (EKF-$\ell_1$) and SMIDAS~\cite{bib24} with learning rate $\eta=5 \cdot 10^{-2}$. Notice that $P_{0|0}, R_k$ and $Q_k$ are chosen by manually tuning EKF-$\ell_1$ and then used for all other approaches, while all the remaining method-specific hyperparameters have been manually tuned to maximize their performance. The reason for choosing a time-varying $\rho_k$ is that fake measurements are usually not accurate initially, so that it is better to start with a higher value of $\frac{1}{\rho_k}$ and then decrease it progressively. In addition, we compare with two offline batch algorithms: NAILM~\cite{bib4} and LBFGS~\cite{bib23}, the latter using the
Python library \texttt{jax-sysid}. Note that all the online algorithms consume the dataset only once (1 epoch), except NAILM and LBFGS that run over 150 and 5000 epochs respectively. The results obtained at the end of the training phase are reported in Table~\ref{tab:ell_1}.
\begin{table}
\addtolength{\tabcolsep}{-4pt}
\begin{center}
\caption{Online learning a static model of~\eqref{eq:static_model} with $\ell_1$ regularization: mean (standard deviation) Loss, Mse, sparsity ratio and execution time obtained over 20 runs.}
\label{tab:ell_1}
\scalebox{0.8}{
\begin{tabular}{l|r|r|r|r} 
  & Loss ($10^{-3}$) & Mse ($10^{-3}$) & Sparsity (\%) & Time [s]\\ [0.5ex] 
  \hline
 LBFGS~\cite{bib23} & $5.40$ ($0.72$) & $1.03$ ($0.19$) & $80.66$ ($5.32$) & $80.51$ ($2.42$) \\ 
 NAILM~\cite{bib4} & $5.24$ ($0.48$) & $1.06$ ($0.15$) & $63.85$ ($5.00$) & $235.41$ ($52.78$) \\ 
 \hline
 EKF-ADMM & $5.99$ ($0.68$) & $1.44$ ($0.17$) & $45.28$ ($4.98$)& $58.27$ ($1.41$) \\
 EKF-ADMMtv & $5.27$ ($0.46$) & $1.29$ ($0.42$) & $57.00$ ($7.95$)& $55.90$ ($1.41$) \\
 online-ADMM~\cite{bib2} & $10.38$ ($1.7$) & $4.68$ ($1.8$) & $3.62$ ($2.21$)& $530.69$ ($29.09$)\\
 EKF-$\ell_1$~\cite{bib1} & $5.47$ ($0.67$) & $1.42$ ($0.26$) & $56.42$ ($7.67$)& $12.46$ ($0.27$)\\
SMIDAS~\cite{bib24} & $89.93$ ($36.3$) & $84.67$ ($36.6$) & $61.71$ ($9.98$)& $4.11$ ($1.11$)\\
\hline
\end{tabular}}
\end{center}
\end{table}
Among the online approaches, EKF-ADMMtv provides the lowest loss: this is also true during the training phase, as shown in Figure~\ref{nn_static_l1_loss}. 
The online learning performance of EKF-ADMM can be also evaluated by looking at the regret function in Figure~\ref{nn_static_l1_regret}, where it is also apparent that, compared to the other approaches, the proposed algorithm improves the solution quality faster.
\subsubsection{Bound constraints}
\begin{figure}[!t]
\centerline{\includegraphics[width=0.94\columnwidth, trim=12 10 30 8, clip]{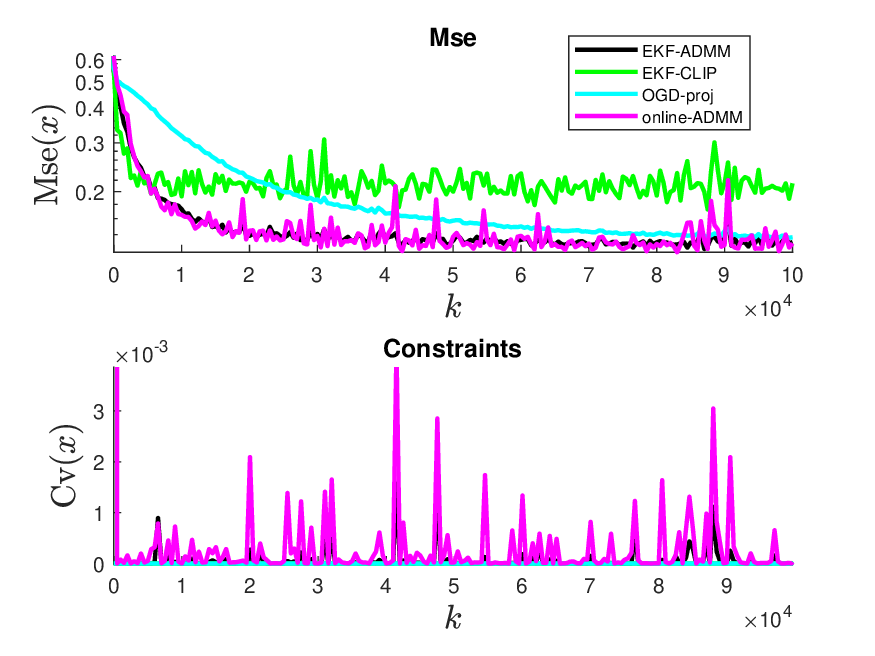}}
\caption{Online learning with bounds: Mse and constraints violation averaged over 20 runs.}
\label{nn_static_BD_loss}
\end{figure}

Let us now repeat the training under the bound constraints imposed by the regularization function $g(x)=0$ if $x \in \mathcal{C}$ and $g(x)=+\infty$ otherwise,  where $\mathcal{C} = \{ x \in \mathbb{R}^{n_x}:\ |x_i|\leq 0.5\}$.  We use the hyper-parameters  $\rho = 1, n_a = 5, Q_k = 10^{-4} I, P_{0|0} = 100 I$, and $R_k = I$. In this example, we also compare with a simple clipping step of the Kalman filter (EKF-CLIP) and an online projected gradient method (OGD-proj)~\cite{bib7} with learning rate $\eta=10^{-3}$. Matrices $P_{0|0}, R_k$ and $Q_k$ are the same as for the $\ell_1$ example and are in common for all the Kalman-like approaches, while all the other hyperparameters  have been manually re-tuned to optimize performance in the bounds constraints example. Results obtained at the end of the training phase are reported in Table~\ref{tab:bounds}. Among the online approaches, considering the final Mse, Cv and execution time, EKF-ADMM provides the best quality solution, while OGD-proj attains similar performance but at a slower pace. Figure~\ref{nn_static_BD_loss} shows the performance of the solution during the training phase.

\begin{table}
\centering
\addtolength{\tabcolsep}{-4pt}
\begin{center}
\caption{Online learning a static model~\eqref{eq:static_model} with bound constraints: mean (standard deviation) Mse, constraints violation, and execution time obtained over 20 runs.}
\label{tab:bounds}
\scalebox{0.9}{
\begin{tabular}{l|r|r|r} 
  & Mse & Cv ($10^{-6}$) & Time [s]\\ [0.5ex] 
 \hline
 LBFGS~\cite{bib23} & $0.122$ ($0.011$) & $0$ ($0$) & $75.87$ ($5.58$)\\ 
 NAILM~\cite{bib4} & $0.137$ ($0.013$) & $0.38$ ($0.71$) & $101.82$ ($3.88$)\\ 
 \hline
 EKF-ADMM & $0.131$ ($0.011$) & $10.76$ ($4.93$) & $70.46$ ($3.45$) \\
 online-ADMM~\cite{bib2} & $0.129$ ($0.010$) & $90.77$ ($59.37$) & $610.73$ ($9.82$) \\
 EKF-CLIP & $0.214$ ($0.048$) & $0$ ($0$) & $11.89$ ($0.12$) \\
 OGD-proj~\cite{bib7} & $0.137$ ($0.013$) & $0$ ($0$) & $2.56$ ($0.15$) \\
 \hline
\end{tabular}}
\end{center}
\end{table}

\subsection{Online learning of a time-varying model}
\begin{figure}[!t]
\centerline{\includegraphics[width=0.94\columnwidth, trim=6 10 30 8, clip]{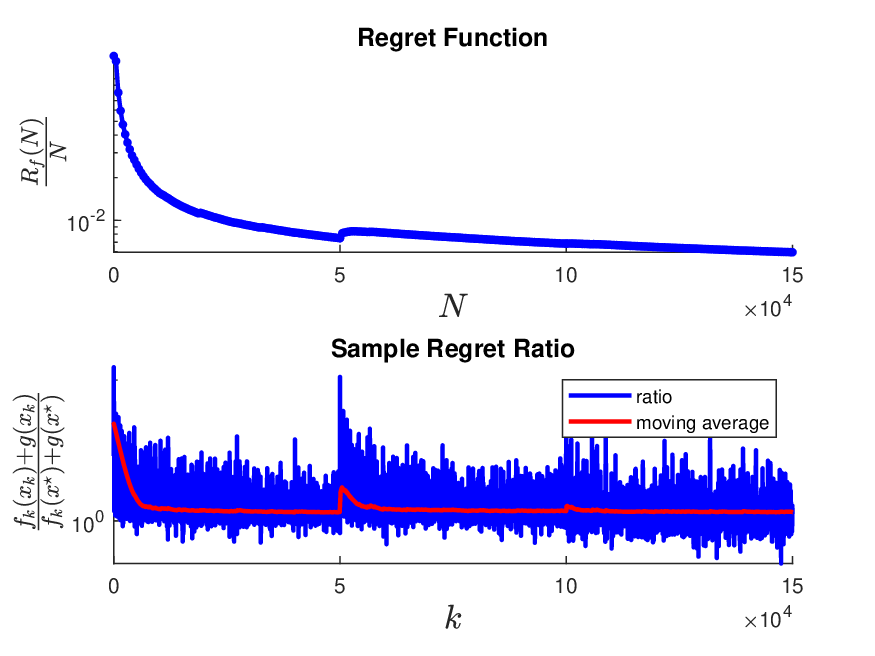}}
\caption{Online learning with $\ell_0$ regularization: regret and sample regret averaged over $20$ runs.}
\label{nn_dynamic_l0_regret2}
\end{figure}

We test now the ability of EKF-ADMM to adapt the same neural network model, under $\ell_0$ regularization, when 
the data-generating system switches as follows: 
$y^m_k =\frac{z_{k,1}^2-e^{\frac{z_{k,2}}{10}}}{3+|z_{k,1}+z_{k,2}|} + r_k$ if $k \leq \frac{N}{3}$, $y^m_k=\frac{z_{k,1}^2-e^{\frac{z_{k,2}}{2}}}{3+|z_{k,1}+z_{k,2}|} + r_k$ if $\frac{N}{3} < k \leq \frac{2 N}{3}$ and $y^m_k = \frac{0.3 \cdot z_{k,1}^2-e^{\frac{z_{k,2}}{2}}}{3+|z_{k,1}+z_{k,2}|} + r_k$ if $\frac{2 N}{3} < k$, with $N = 1.5 \cdot 10^{5}$, $z_{k,i} \sim \mathcal{U}_{[-10,10]}$ and $r_k\sim {\mathcal{N}}(0,10^{-2})$. We evaluate the regret function $R_{f}(N) = \sum_{k=0}^{N-1} (f_k(x_k)+g(x_k)) - \min_{z_1,z_2,z_3}\sum_{i=1}^{3} r_i(z_i)$, with $r_i(z_i) = \sum_{k=(i-1)\frac{N}{3}}^{i\frac{N}{3}}(f_k(z_i)+g(z_i))$, where $\{ x_k\}_{k=0}^{N-1}$ is the sequence generated by Algorithm \ref{alg1}. The regularization term is $g(x) = \lambda \norm{w}_0$, with $\lambda = 10^{-4}$, and we use the EKF-ADMM hyper-parameters $\rho = 10^3 \cdot \lambda$, $p_a = 1$,  $Q_k = 10^{-4} I$, and $P_{0|0} = 100 I$, which were manually tuned by optimizing the EKF performance. Since the model is now time-varying, we will also use an EKF implementation with forgetting factor $\alpha = 0.9$ \cite{bib3}. The resulting regret function is shown in Figure~\ref{nn_dynamic_l0_regret2}. It is apparent that EKF-ADMM can effectively track changes of the underlying data-generating system. 

We have noticed that the choice of $\rho$ is crucial to obtain good adaptation performance. While for some values of $\rho$ the convergence is slow and not satisfactory, with limited effort, in all the reported examples, we could find appropriate values that gave good performance. We remark that, in the general nonlinear case, this might not always be the case, as both the EKF and ADMM are not guaranteed to converge, which might complicate tuning $\rho$. This could limit the use of ADMM iterations within the EKF framework in certain challenging applications.

\section{Conclusions}
\label{sec:conclusions}
We have proposed a novel algorithm for online learning of nonlinear parametric models under non-smooth regularization using a combination of EKF and ADMM, for which we derived a sublinear regret bound for the convex linear time-varying case. The approach is computationally cheap and is suitable for factorized or square-root implementations that can make it numerically robust, and is therefore very appealing for embedded applications of adaptive control, such as adaptive model predictive control. The effectiveness of the approach has been evaluated in three numerical examples. Future investigations will focus on extending the approach to the recursive identification of parametric nonlinear state-space dynamical models from input/output data under non-smooth regularization, in which both the hidden states and the parameters are jointly estimated.
\bibliographystyle{plain}
\bibliography{bibliography}

@article{bib1,
  author = {A. Bemporad},
  title = {Recurrent Neural Network Training with Convex Loss and Regularization Functions by Extended Kalman Filtering},
  journal = {IEEE Transactions on Automatic Control},
  volume = {68},
  number = {1},
  pages = {5661-5668},
  year = {2021}
}

@inproceedings{bib2,
  author = {H. Wang and A. Banjaree},
  title = {Online alternating direction method},
  booktitle = {Proc. 29th International Conference on Machine Learning},
  address = {Edinburgh, Scotland, UK},
  pages = {1699--1706},
  year = {2012}
}

@inproceedings{bib3,
  author = {A. Abulikemu and L. Changliu},
  title = {Robust Online Model Adaptation by Extended {Kalman} Filter with Exponential Moving Average and Dynamic Multi-Epoch Strategy},
  booktitle = {Proc. 2nd Conference on Learning for Dynamics and Control},
  volume = {120},
  pages = {65--74},
  year = {2020}
}

@article{bib4,
  author = {A. Bemporad},
  title = {Training recurrent neural networks by sequential least squares and the alternating direction method of multiplier},
  journal = {Automatica},
  volume = {156},
  number = {1},
  pages = {111183},
  year = {2023}
}

@article{bib5,
  author = {S. Boyd and N. Parikh and E. Chu and B. Peleato and J. Eckstein},
  title = {Distributed Optimization and Statistical Learning via the Alternating Direction Method of Multipliers},
  journal = {Foundations and Trends in Machine Learning},
  volume = {3},
  number = {1},
  pages = {1-122},
  year = {2011}
}

@article{bib7,
  author = {S.C.H. Hoi and D. Sahoo and J. Lu and P. Zhao},
  title = {Online learning: A comprehensive survey},
  journal = {Neurocomputing},
  volume = {459},
  pages = {249-289},
  year = {2021}
}

@article{bib8,
  author = {G. Pillonetto and A. Aravkin and D. Gedon and L. Ljung and A. H. Ribeiro and T. B. Schön},
  title = {Deep networks for system identification: A survey},
  journal = {Automatica},
  volume = {171},
  pages = {111907},
  year = {2025}
}

@article{bib9,
  author = {J. Schoukens and L. Ljung},
  title = {Nonlinear System Identification: A User-Oriented Road Map},
  journal = {IEEE Control Systems},
  volume = {39},
  pages = {28-99},
  year = {2019}
}

@article{bib10,
    author = {Vincent A. Akpan and George D. Hassapis},
    title = {Nonlinear model identification and adaptive model predictive control using neural networks},
    journal = {ISA Transactions},
    volume = {50},
    number = {2},
    pages = {177-194},
    year = {2011}
}

@article{bib13,
  author = {H. Jeffrey and R. Preston and W. Jeremy},
  title = {A Fresh Look at the {Kalman} Filter},
  journal = {SIAM Review},
  volume = {54},
  number = {4},
  pages = {801-823},
  year = {2012}
}

@article{bib14,
  author = {R.E. Kalman},
  title = {A New Approach to Linear Filtering and Prediction Problems},
  journal = {ASME. J. Basic Eng},
  volume = {82},
  number = {1},
  pages = {35-45},
  year = {1960}
}

@article{bib15,
  author = {N. Parikh and S. Boyd},
  title = {Proximal Algorithms},
  journal = {Foundations and Trends in Optimization},
  volume = {1},
  number = {3},
  pages = {123-231},
  year = {2013}
}

@inproceedings{bib16,
  author = {C.L. Thornton and G.J. Bierman},
  title = {{Gram-Schmidt} algorithms for covariance propagation},
  booktitle = {IEEE Conference on Decision and Control},
  pages = {489-498},
  year = {1975}
}

@article{bib18,
  author = {Y. Zhang and Z. Xiao and J. Wu and L. Zhang},
  title = {Online Alternating Direction Method of Multipliers for Online Composite Optimization},
  journal = {arXiv preprint arXiv:1904.02862},
  year = {2024}
}

@article{bib20,
  title={A Survey of Algorithms and Analysis for Adaptive Online Learning},
  author={H.B. McMahan},
  journal={Journal of Machine Learning Research},
  pages={1--50},
  year={2017},
  volume={18},
}

@article{bib22,
    author = {J. Ranstam and J.A. Cook},
    title = {{LASSO} regression},
    journal = {British Journal of Surgery},
    volume = {105},
    number = {10},
    pages = {1348-1348},
    year = {2018}
}

@article{bib23,
    author={A. Bemporad},
    title={An {L-BFGS-B} approach for linear and nonlinear system identification under $\ell_1$ and group-Lasso regularization},
    journal = {IEEE Transactions on Automatic Control},
    note = {in press.},
    year=2025
}

@inproceedings{bib24,
author = {Shalev-Shwartz, Shai and Tewari, Ambuj},
title = {Stochastic methods for $\ell_1$ regularized loss minimization},
year = {2009},
booktitle = {Proceedings of the 26th Annual International Conference on Machine Learning},
pages = {929–936},
numpages = {8},
location = {Montreal, Quebec, Canada},
series = {ICML '09}
}

@inbook{bib25,
author = {Sara van de Geer},
title = {$\ell_1$-regularization in High-dimensional Statistical Models},
booktitle = {Proceedings of the International Congress of Mathematicians 2010 (ICM 2010)},
chapter = {},
pages = {2351-2369},
}

@article{bib26,
title = {Group sparse regularization for deep neural networks},
journal = {Neurocomputing},
volume = {241},
pages = {81-89},
year = {2017},
author = {Simone Scardapane and Danilo Comminiello and Amir Hussain and Aurelio Uncini},
}

@Article{AGHRD19,
  author = {J.A.E. Andersson and J. Gillis and G. Horn
            and J.B. Rawlings and M. Diehl},
  title = {{CasADi} -- {A} software framework for nonlinear optimization
           and optimal control},
  journal = {Mathematical Programming Computation},
  volume = {11},
  number = {1},
  pages = {1--36},
  year = {2019},
}
\appendix
\section{Proof of Theorem 1}
\label{app:Theorem1}
Starting from $R_f(N)$, since $x_{k+1}, \nu_{k+1}$ are the optimal solutions of the first two optimization problems in Algorithm \ref{alg3} and since $w_{k} = w_{k+1}-\rho(x_{k+1}-\nu_{k+1})$ we have that $\nabla f_{k}(x_{k+1})=  -(w_{k+1}-\rho(\nu_k - \nu_{k+1})) - \eta(P_{k}^{-1}x_{k+1}-P_{k}^{-1}x_{k})$ and $w_{k+1} \in  \partial g(\nu_{k+1})$, where $\partial g$ is the subgradient of $g$. Due to the convexity of $f_k(\cdot)$ and $g(\cdot)$, $f_k(x_{k+1})-f_k(x^\star)  \leq \nabla f_{k}(x_{k+1})^T(x_{k+1}-x^\star)=  -w_{k+1}^T(x_{k+1}-\nu^\star) + \frac{\rho}{2}(\norm{\nu^\star -\nu_k}^2-\norm{\nu^\star -\nu_{k+1}}^2+ \norm{x_{k+1}-\nu_{k+1}}^2-\norm{x_{k+1}-\nu_{k}}^2)+\frac{\eta}{2} (\norm{x^\star-x_k}^2_{P_k^{-1}} -$ $\norm{x^\star-x_{k+1}}^2_{P_k^{-1}}-\norm{x_{k+1}-x_k}^2_{P_k^{-1}})$, where the equality derives from a succession of square completions and by recalling that $x^\star=\nu^\star$, and $g(\nu_{k+1})-g(\nu^\star) \leq w_{k+1}^T(\nu_{k+1}-\nu^\star)$.
Summing the two inequalities together and noticing that $-w_{k+1}^T(x_{k+1}-\nu_{k+1})+\frac{\rho}{2}\lVert x_{k+1}-\nu_{k+1}\rVert^2 = \frac{1}{2\rho}(\norm{w_k}^2-\norm{w_{k+1}}^2)$, we obtain: 
\begin{equation}
\begin{aligned}
& f_k(x_{k+1})+g(\nu_{k+1})-(f_k(x^\star)+g(\nu^\star))  \leq \\ & \leq \frac{1}{2\rho}(\norm{w_k}^2-\norm{w_{k+1}}^2) - \frac{\rho}{2}\norm{x_{k+1}-\nu_k}^2 + \\
& \frac{\rho}{2} (\norm{\nu^\star - \nu_k}^2-\norm{\nu^\star - \nu_{k+1}}^2) +  \frac{\eta}{2} (\norm{x^\star-x_k}^2_{P_k^{-1}}\\
& -\norm{x^\star-x_{k+1}}^2_{P_k^{-1}}-\norm{x_{k+1}-x_k}^2_{P_k^{-1}}) \\
\end{aligned}
\label{thmeqn}
\end{equation}
Considering that $f_k(x_k)-f_k(x_{k+1}) \leq \nabla f_k(x_k)^T (x_k-x_{k+1}) \leq \frac{1}{2\alpha \eta} \norm{\nabla f_k(x_k)}^2+ \frac{\alpha \eta}{2}\norm{x_k - x_{k+1}}^2$, where the second inequality is due to Fenchel-Young's inequality, and considering Assumption~A1.a of the theorem, we have that $f_k(x_k)+g(\nu_{k+1})-(f_k(x^\star)+g(\nu^\star)) \leq \frac{1}{2\rho} (\norm{w_k}^2- \norm{w_{k+1}}^2) -\frac{\rho}{2}\norm{x_{k+1}-\nu_k}^2+ \frac{\rho}{2}(\norm{\nu^\star - \nu_k}^2-\norm{\nu^\star - \nu_{k+1}}^2)+   \frac{1}{2\alpha \eta} \norm{\nabla f_k(x_k)}^2+\frac{\eta}{2} (\norm{x^\star - x_k}_{P_{k}^{-1}}^2-\norm{x^\star - x_{k+1}}_{P_{k}^{-1}}^2)$. Summing from 0 to $N-1$ and considering Assumption~A3 we get
\[
\begin{aligned}
& R_f(N) =  \sum_{k=0}^{N-1}(f_k(x_k)+g(\nu_{k+1})-(f_k(x^\star)+g(\nu^\star)))\\
& + g(\nu_0)-g(\nu_N) \leq   \frac{1}{2\rho} (\norm{w_0}^2 - \norm{w_N}^2)+\frac{\rho}{2}(\norm{\nu^\star - \nu_0}^2\\
&-\norm{\nu^\star - \nu_{N}}^2) + \frac{1}{2\alpha \eta} \sum_{k=0}^{N-1} \norm{\nabla f_k(x_k)}^2 + \frac{\eta}{2} \norm{x^\star-x_0}^2_{P_0^{-1}}\\
 & + \frac{\eta}{2} \sum_{k=1}^{N-1} \norm{x^\star - x_{k}}^2_{(P_{k}^{-1}-P_{k-1}^{-1})} \\
\end{aligned}
\]
and, therefore, $R_f(N) \leq \frac{\rho}{2}\norm{\nu^\star}^2$ $+\frac{1}{2\alpha \eta} \sum_{k=0}^{N-1} \norm{\nabla f_k(x_k)}^2 + \frac{\eta}{2} \norm{x^\star}^2_{P_0^{-1}} + \frac{\eta}{2} \sum_{k=1}^{N-1} \norm{x^\star - x_{k}}^2_{(P_{k}^{-1}-P_{k-1}^{-1})}$. Because of Assumption A2, $\frac{1}{2}\sum_{k=1}^{N-1} \norm{x^\star - x_{k}}^2_{(P_{k}^{-1}-P_{k-1}^{-1})} = \frac{1}{2}\sum_{k=1}^{k_n} \norm{x^\star - x_{k}}^2_{(P_{k}^{-1}-P_{k-1}^{-1})} \leq M_{k_n}$, and taking into account Assumptions~A1.b and~A1.c we get $R_f(N)\leq \frac{\rho D_{\nu}}{2} + \frac{N G_f^2}{2\alpha \eta} + \eta (D_x^2 + M_{k_n})$. Setting $\eta \defeq \frac{G_f \sqrt{N}}{D_x \sqrt{2\alpha}}$ and $ \rho \defeq \sqrt{N}$, we get the sublinear regret bound $R_f(N)\leq \frac{\sqrt{N} D_{\nu}}{2} + \frac{G_f D_x\sqrt{N} }{\sqrt{2\alpha}} + \frac{G_f \sqrt{N}(D_x^2 + M_{k_n})}{D_x \sqrt{2\alpha}}$. Considering now $R_c(N)$, we can rearrange~\eqref{thmeqn} and consider Assumption~A1.d,  $\norm{x_{k+1} - \nu_k}^2 \leq  \frac{2F}{\rho}+ \frac{1}{\rho^2} (\norm{w_k}^2 - \norm{w_{k+1}}^2) +  (\norm{\nu^\star - \nu_k}^2-\norm{\nu^\star - \nu_{k+1}}^2) + +  \frac{\eta}{\rho}\left(\norm{x^\star - x_k}_{P_{k}^{-1}}^2-\norm{x^\star - x_{k+1}}_{P_{k}^{-1}}^2 - \norm{x_{k+1} - x_{k}}_{P_{k}^{-1}}^2\right).$ Summing from 0 to $N-1$, we get: 
\[ \begin{aligned}
&R_c(N) =  \sum_{k=0}^{N-1}\norm{x_{k+1} - \nu_k}^2 \leq  \frac{2FN}{\rho} + \norm{\nu^\star}^2 \\ 
 &+ \frac{\eta}{\rho} \bigg{(} \norm{x^\star}^2_{P_0^{-1}}+  \sum_{k=1}^{N-1} \norm{x^\star - x_{k}}^2_{(P_{k}^{-1}-P_{k-1}^{-1})}  \bigg{)}.
\end{aligned}
\]
Considering Assumptions A1.c and A2, we have $R_c(N) \leq \frac{2FN}{\rho} + D_{\nu} + \frac{2\eta}{\rho} (D_x^2+M_{k_n})$ and setting $\eta \defeq \frac{G_f \sqrt{N}}{D_x \sqrt{2\alpha}}$ and $ \rho \defeq \sqrt{N}$ we finally get $R_c(N) \leq 2F\sqrt{N} + D_{\nu} + \frac{2G_{f}}{D_x \sqrt{2\alpha}} (D_x^2+M_{k_n})$.
\hfill$\Box$
\end{document}